       
\def\bs{\bigskip}

\def\ni{\noindent}
\def\cl{\centerline}

\def\ref#1#2#3#4{#1\ {\it#2\ }{\bf#3\ }#4\par}
\def\refb#1#2#3{#1\ {\it#2\ }#3\par}
\def\CQG{Class.\ Qu.\ Grav.}
\def\GRG{Gen.\ Rel.\ Grav.}
\def\PL{Phys.\ Lett.}
\def\PR{Phys.\ Rev.}
\def\PRS{Proc.\ R.\ Soc.\ Lond.}

\def\d{\delta}
\def\e{\varepsilon}
\def\p{\partial}
\def\half{{\textstyle{1\over2}}}
\def\sixth{{\textstyle{1\over6}}}

\magnification=\magstep1

\cl{\bf Comment on ``Failure of standard conservation laws}
\cl{\bf at a classical change of signature''}
\bs\cl{\bf Sean A. Hayward}
\cl{Department of Physics,
Kyoto University,
Kyoto 606-01,
Japan}
\bs\ni
{\bf Abstract.}
Hellaby \& Dray have recently claimed that 
matter conservation fails under a change of signature,
compounding earlier claims that 
the standard junction conditions for signature change are unnecessary.
In fact, if the field equations are satisfied,
then the junction conditions and the conservation equations are satisfied.
The failure is rather that the authors did not make sense of 
the field equations and conservation equations,
which are singular at a change of signature.
\bs\ni
In quantum cosmology, 
there are so-called real tunnelling solutions 
to the Wheeler-deWitt equation [1--8]
which may be interpreted in a purely classical way 
as solutions to the classical field equations 
in which the metric changes signature from Lorentzian to Riemannian.
In the quantum cosmology approach, 
the momentum fields are real in the Lorentzian region 
and imaginary in the Riemannian region,
and so vanish at the junction.
These junction conditions may also be derived 
by various purely classical methods [8--17].
Several authors omitted these junction conditions [18--26]
and some subsequently claimed explicitly that they are unnecessary [20--22,26].
Some of the mistakes behind these claims 
have been pointed out already [9,10,12,14].
Unfortunately, some of these authors still insist 
that the junction conditions are not required by the field equations.
The most recently published example is the paper of Hellaby \& Dray [26],
which purports to show that 
matter conservation need not hold under a change of signature.
It is easily seen that this is due to a failure 
to satisfy the field equations at the junction, i.e.\ the junction conditions.

It is shown here that in the approach favoured by Hellaby \& Dray,
the field equations are (i) well defined in a distributional sense,
and (ii) not satisfied unless the momentum fields vanish at the junction.
When the field equations are satisfied,
the conservation equations are also well defined and satisfied.

For simplicity, consider homogeneous isotropic cosmologies,
for which the line-element may be written as
$$ds^2=-N^{-1}dt^2+a^2d\Sigma^2\eqno(1)$$
where $d\Sigma^2$ refers to a constant-curvature space,
and the scale factor $a$ and inverse squared lapse $N$ are functions of $t$.
For concreteness, take the matter model to be a scalar field $\phi$,
a real function of $t$,
with potential $V(\phi)$.
The Einstein-Klein-Gordon equations may be written
$$\eqalignno{
&Na''+\half N'a'=-\sixth Na(\phi')^2+\sixth aV&(2a)\cr
&N\phi''+\half N'\phi'=-3Na^{-1}a'\phi'-\p V/\p\phi&(2b)\cr
&0=12(a')^2N-(a\phi')^2N+2(6k-a^2V)&(2c)\cr}$$
where the prime denotes $\p/\p t$,
and $k=-1$, 0, 1 labels the hyperbolic, flat and spherical cases respectively.
All quantities are real.
A change of signature occurs if $N$ changes sign,
the junction surface being given by $N=0$.
Note that the evolution equations (2a,b) are singular at a change of signature:
at $N=0$ they do not determine $(a'',\phi'')$ but constrain $(a',\phi')$.
Hellaby \& Dray consider taking $N=\e$,
where $\e$ is the sign of $t$.
Since $\e'=2\d$,
where $\d$ is the Dirac distribution [27,28] 
with support at the junction $t=0$,
the field equations are well defined in a distributional sense:
$$\eqalignno{
&\e a''+\d a'=-\sixth\e a(\phi')^2+\sixth aV&(3a)\cr
&\e\phi''+\d\phi'=-3\e a^{-1}a'\phi'-\p V/\p\phi&(3b)\cr
&0=12(a')^2\e-(a\phi')^2\e+2(6k-a^2V).&(3c)\cr}$$
The singular parts of the field equations are the junction conditions:
$$\eqalignno{
&a'\d=0&(4a)\cr
&\phi'\d=0.&(4b)\cr}$$
In other words, $a'$ and $\phi'$ vanish at the junction.
The conservation equation 
is the time derivative of the constraint equation (3c),
which is well defined in a distributional sense
and is identically satisfied as a consequence of the other field equations.

Hellaby \& Dray miss these conditions 
because they do not write the field equations explicitly.
Instead, they calculate jumps in various quantities,
without checking whether these quantities are well defined at the junction.
In particular, they use the second fundamental form,
which contains $N$ to the power $1/2$, as do momentum fields generally.
These singular momentum fields can be properly defined in various ways [8--17],
but Hellaby \& Dray do not do so.
Hellaby \& Dray have neither well defined fields
nor well defined field equations,
and so they fail to obtain the junction conditions.
Identical or similar problems occur in the previous papers [18--25],
as pointed out previously [9,10,12,14].

Hellaby \& Dray try to justify their claims 
by appealing to the Darmois conditions,
continuity of the first and second fundamental forms of the junction.
These conditions were formulated for non-degenerate metrics,
where they suffice to prevent distributional terms 
in the vacuum Einstein equations [29--31].
They do not suffice to prevent distributional terms 
in the vacuum Einstein equations across a change of signature, as shown above.
The underlying reason is that 
the field equations are singular at a change of signature, 
so that the usual ad hoc matching rules do not work.
The singular nature of the field equations at a change of signature 
deserves emphasis,
since much of the confusion seems to stem from 
naive application of rules learned for non-singular equations.

Hellaby \& Dray argue that 
their approach involves some minimal set of assumptions,
and that other (unspecified) approaches are more restrictive 
and therefore less satisfactory.
However, their assumptions are so minimal that 
they do not suffice to make sense of the field equations.
In all the suggested approaches, including that favoured by Hellaby \& Dray,
the junction conditions follow from the field equations,
once the latter are well defined [1--17].
The junction conditions are the field equations at the junction.

The principal claim of Hellaby \& Dray,
that conservation laws fail at a change of signature, 
is incorrect.\footnote\dag{This should have been clear from the outset:
formal calculations yield the contracted Bianchi identity 
$\nabla^aG_{ab}=0$ as usual,
the Einstein equations $G_{ab}=T_{ab}$ 
then formally yielding matter conservation $\nabla^aT_{ab}=0$.}
The failure is rather that 
these authors did not make sense of the conservation equations,
nor of the field equations themselves.
Nevertheless,
the field equations and the consequent conservation equations are well defined.
The field equations are satisfied only if 
the standard junction conditions are satisfied,
in which case the conservation equations are also satisfied.
This justifies the usual approach in quantum cosmology.
It also confirms that matter conservation has its usual physical status.
\bs\ni
For lengthy discussions or correspondence over the last three years 
I acknowledge
T.~Dray, G.~Ellis, M.~Kriele, R.~Tucker and several anonymous referees.
\bs
\begingroup
\parindent=0pt\everypar={\global\hangindent=20pt\hangafter=1}\par
{\bf References}\par
\ref{[1] Vilenkin A 1982}\PL{117B}{25}
\ref{[2] Halliwell J J \& Hartle J B 1990}\PR{D41}{1815}
\ref{[3] Gibbons G W \& Hartle J B 1990}\PR{D42}{2458}
\ref{[4] Fujiwara Y, Higuchi S, Hosoya A, Mishima T \& Siino M 1991}
\PR{D44}{1756}
\ref{[5] Hawking S W, Laflamme R \& Lyons G W 1993}\PR{D47}{5342}
\refb{[6] Vilenkin A 1993}{Quantum cosmology}{(gr-qc/9302016)}
\refb{[7] Barvinsky A O \& Kamenshchik A Y 1993}{Tunnelling geometries I}
{(gr-qc/9311022)}
\ref{[8] Carlip S 1993}\CQG{10}{1057}
\ref{[9] Hayward S A 1992}\CQG9{1851; erratum 2453}
\refb{[10] Hayward S A 1992}
{Comment on ``Change of signature in classical relativity''}{(unpublished)}
\ref{[11] Hayward S A 1993}\CQG{10}{L7}
\refb{[12] Hayward S A 1994}{Junction conditions for signature change}
{(gr-qc/9303034)}
\ref{[13] Kossowski M \& Kriele M 1993}\CQG{10}{1157}
\ref{[14] Kossowski M \& Kriele M 1993}\CQG{10}{2363}
\ref{[15] Kossowski M \& Kriele M 1994}\PRS{A444}{297}
\ref{[16] Kossowski M \& Kriele M 1994}
{The Einstein equation for signature-type-changing space-times, \PRS}A
{(in press)}
\refb{[17] Kossowski M \& Kriele M 1994}{Characteristic classes for 
transverse type-changing pseudo-Riemannian metrics with smooth curvature}
{(preprint)}
\ref{[18] Dray T, Manogue C A \& Tucker R W 1991}\GRG{23}{967}
\refb{[19] Dray T, Manogue C A \& Tucker R W 1992}
{The effect of signature change on scalar field propagation}{(unpublished)}
\ref{[20] Dray T, Manogue C A \& Tucker R W 1993}\PR{D48}{2587}
\ref{[21] Ellis G F R 1992}\GRG{24}{1047}
\ref{[22] Ellis G, Sumeruk A, Coule D \& Hellaby C 1992}\CQG{9}{1535}
\ref{[23] Romano J D 1993}\PR{D47}{4328}
\ref{[24] Kerner R \& Martin J 1993}\CQG{10}{2111}
\ref{[25] Martin J 1994}\PR{D49}{5086}
\ref{[26] Hellaby C \& Dray T 1994}\PR{D49}{5096}
\refb{[27] Schwartz L 1950}{Th\'eorie des Distributions}{(Paris: Hermann)}
\refb{[28] Gel'fand I M \& Shilov G E 1964}{Generalized Functions}
{(New York: Academic Press)}
\ref{[29] Israel W 1966}{Nuovo Cimento}{B44}{1; erratum {\bf B48} 463}
\ref{[30] Clarke C J S \& Dray T 1987}\CQG4{265}
\ref{[31] Barrab\`es C 1989}\CQG6{581}
\endgroup
\bye